\documentclass[12pt,pra,aps,amssymb,amsfonts,amsmath,tightenlines]{revtex4}
\usepackage{dsfont}
\usepackage{graphicx}
\usepackage{amssymb,amsfonts,amsthm}
\usepackage{color}
\usepackage{verbatim}
\usepackage[normalem]{ulem}
\usepackage{epstopdf}

\newcommand{\half}{\mbox{$\textstyle \frac{1}{2}$}}

\newcommand{\re}{\mbox{$\rm e$}}

\newcommand{\rd}{\mbox{$\rm d$}}
\newtheorem{prop}{Proposition}

\begin{document}

\title{Rational Term Structure Models with \\ Geometric L\'evy Martingales}

\author{Dorje C.~Brody$^*$, Lane~P.~Hughston$^\dagger$, and Ewan~Mackie$^\dagger$}

\affiliation{$^*$Mathematical Sciences, Brunel University, Uxbridge UB8 3PH, UK \\ 
$^\dagger$Department of Mathematics, Imperial College London,
London SW7 2BZ, UK \\
$^\ddagger$Imperial College Business School, London SW7 2BZ, UK}

\date{\today}

\begin{abstract}
In the ``positive interest" models of Flesaker-Hughston,
the nominal discount bond system is determined by a
one-parameter family of positive martingales. In the 
present paper we extend this analysis to include a variety 
of distributions for the martingale family, parameterised by 
a function that determines the behaviour of the market risk 
premium. These distributions include jump and diffusion 
characteristics that generate various properties for discount 
bond returns. For example, one can generate skewness and 
excess kurtosis in the bond returns by choosing the martingale 
family to be given by (a) exponential gamma processes, or (b) 
exponential variance gamma processes. The models are 
``rational" in the sense that the discount bond price is given 
by a ratio of weighted sums of positive martingales. Our 
findings lead to semi-analytical formulae for the prices of 
options on discount bonds. A number of general 
results concerning L\'evy interest rate models are 
presented as well.
\end{abstract}

\maketitle

\section{Interest rate models: the volatility approach}

\noindent From a modern perspective there are two main approaches
to the modelling of interest rates. These are the ``volatility approach" and 
the ``pricing kernel approach". Both have been investigated extensively 
in the case of a Brownian filtration, but rather less so in the general situation 
when bond prices are allowed to jump. The main purpose of this paper is to 
present some new term structure models admitting, jumps based on 
parametric families of geometric L\'evy martingales; but in doing so we shall 
take the opportunity to make some observations about the present state of 
interest rate modelling, both in the Brownian setting and the more general 
L\'evy-Ito setting.

The volatility approach, in the case of a Brownian market filtration, is at 
present perhaps the most important and widely implemented  interest rate 
modelling methodology, and it is the best understood. The celebrated HJM 
framework (Heath \textit{et al}.~1992) belongs to this category, as does also 
the so-called Libor market model in its various manifestations (see e.g. 
Rebonato 2002, Musiela \& Rutkowski 2005, and references cited therein), 
and a number of the ``classical" short-rate models can be reformulated in this 
way as well. The volatility approach thus deserves special attention. We shall 
discuss the Brownian case first, and then its rather less well understood 
extension to the jump category.

The setup is as follows. The discount bond volatility, the market
price of risk, and the initial yield curve, constitute the ``primitive
data" of the model. We have a probability space $(\Omega,{\mathcal F},
{\mathbb P})$ equipped with the augmented filtration
$\{{\mathcal F}_t\}_{t\geq0}$ associated with a Brownian motion
$\{W_t\}_{t\geq0}$ of one or more dimensions. Here and in what follows
${\mathbb P}$ denotes the ``physical'' measure; most of the discussion
will focus on  ${\mathbb P}$ rather than on alternative measures.
The interest rate markets are represented by a system of unit-principal
discount bonds $\{P_{tT}\}_{T\geq t\geq0}$ and a unit-initialised
money-market account $\{B_t\}_{t\geq0}$ satisfying a system of
dynamical equations of the following form:
\begin{eqnarray}\label{eq:1}
{\rd} P_{tT} = r_t P_{tT} {\rd} t + \Omega_{tT} P_{tT} ({\rd} W_t +
\lambda_t{\rd} t),  \quad
{\rd} B_{t} = r_t  B_{t} \, {\rd} t.
\end{eqnarray}
Here $P_{tT}$ is the random value at time $t$ of the bond  that matures at
time $T$, $\{r_t\}_{t\geq0}$ is the short rate, $\{\Omega_{tT}\}_{T\geq t\geq0}$
is the bond volatility, and $\{\lambda_t\}_{t\geq0}$ is the market price of risk.
In the multi-dimensional setting, $\{\Omega_{tT}\}$, $\{\lambda_t\}$, and
$\{W_t\}$ are vector-valued processes, and in equation (1) there is an implied
``dot product" between $\{\Omega_{tT}\}$ and ${\rd}W_t$, and also between
$\Omega_{tT}$ and $\lambda_t$. We shall assume that the family of processes 
$\{P_{tT}\}$ is differentiable with respect to $T$ in a suitable sense. We also 
assume that the family of processes $\{\Omega_{tT}\}$ is differentiable with
respect to $T$ in a suitable sense, and that for any fixed $T$ it holds that 
$\lim_{t\to T} \Omega_{tT}=0$. For some versions of the 
theory (such as the original formulation of HJM) one imposes a finite time 
horizon over which the model is defined; whereas for other versions the time 
horizon is infinite. We mostly consider the latter case here, and we assume 
that $\lim_{T\to \infty}P_{0T}=0$.

Given the initial bond prices $\{P_{0T}\}_{T \geq 0}$, we find, under suitable
conditions, that the solution for the discount bond system is
\begin{eqnarray}\label{eq:3}
P_{tT} = P_{0T}B_{t} \exp \left( \int_0^t \Omega_{sT} \, ({\rd} W_s
+ \lambda_s \, {\rd} s) - \half \int_0^t \Omega^2_{sT}\, {\rd} s
\right),
\end{eqnarray}
and that for the money-market account we have
\begin{eqnarray}\label{money market account}
B_{t} =  \exp   \left( \int_0^t r_s \, {\rd} s \right).
\end{eqnarray}
We require, in particular,  that the conditions
\begin{eqnarray}
\int_0^t  \left| \Omega_{sT} \, \lambda_s \right| \,  {\rd} s  < \infty,  \quad
\int_0^t \Omega^2_{sT}\, {\rd} s < \infty,  \quad
\int_0^t  \left| r_s \right| \, {\rd} s < \infty
\end{eqnarray}
hold almost surely for all $t>0$. If we set  $\lim_{t\to T}P_{tT}=1$ for all 
$T\geq0$, we can invert equation (\ref{eq:3}) to obtain
\begin{eqnarray}\label{eq:4}
B_t = (P_{0t})^{-1} \exp \left( - \int_0^t \Omega_{st} \, ({\rd} W_s +
\lambda_s \, {\rd} s ) + \half \int_0^t \Omega^2_{st} \, {\rd} s
\right).
\end{eqnarray}
For the short rate $r_t$ we then have
\begin{eqnarray}
r_t = - \partial_t \ln P_{0t} + \int_0^t \Omega_{st} \partial_t
\Omega_{st} \, {\rd} s - \int_0^t \partial_t \Omega_{st} \, ({\rd} W_s
+ \lambda_s {\rd} s),
\end{eqnarray}
where $\partial_t$ denotes differentiation with respect to $t$. Putting these
ingredients together we are led to the following expression
for the bond prices:
\begin{eqnarray}
P_{tT} = P_{0tT}\,
\frac{\exp \left( \int_0^t \Omega_{sT} \, ({\rd} W_s + \lambda_s \, {\rd} s) -
\frac{1}{2} \int_0^t \Omega ^2_{sT} \, {\rd} s \right)} {\exp \left( \int_0^t
\Omega_{st} \, ({\rd} W_s + \lambda_s\, {\rd} s) - \frac{1}{2} \int_0^t
\Omega ^2_{st} {\rd} s \right)}.
\end{eqnarray}
Here $P_{0tT}=P_{0T}/P_{0t}$ denotes the $t$-forward price made at time
$0$ for a $T$-maturity bond. We see that $\{B_t\}$  and $\{P_{tT}\}$ are
determined by the specification of the volatility $\{\Omega_{tT}\}$, the market
price of risk $\{\lambda_t\}$, and the initial bond prices $\{P_{0t}\}$. This is
the sense in which these are the primitive data of the model. In particular,
there is no need to model the short rate as such separately---it is a ``derived"
quantity in the volatility approach. Two further conditions are required,
namely, that the processes $ \{ \Lambda_t \} $ and  $ \{ \bar P_{tT} \} $
defined by
\begin{eqnarray}
\Lambda_t =  \exp \left(- \int_0^t \lambda_s
\, {\rd} W_s - \half \int_0^t \lambda_s^2 \, {\rd} s \right).
\end{eqnarray}
for $t \geq 0$, and
\begin{eqnarray}
\bar P_{tT} = P_{0T} \exp \left( \int_0^t (\Omega_{sT}-\lambda_s)
\, {\rd} W_s - \half \int_0^t (\Omega_{sT}-\lambda_s)^2 \, {\rd} s \right).
\end{eqnarray}
for $0 \leq t \leq T$, are martingales (rather than merely local martingales).
These conditions are needed if we are to make economic sense of the models
and to put them into practice.

But how do we put the volatility approach into practice? As long as we are
primarily interested in pricing and hedging, but less so (at least from a
modelling perspective) in portfolio allocation, scenario analysis,  and
forecasting, then a transformation to the risk-neutral measure $\mathbb Q$
has the effect of removing the market price of risk from the equations; and
the problem of modelling the evolution of the term structure of interest rates
is transformed into the problem of modelling $\{\Omega_{tT}\}$ under
$\mathbb Q$. In the methodology that has been adapted by practitioners
the idea is that we specify $\{\Omega_{tT}\}$ exogenously, under 
$\mathbb Q$ or under an appropriate set of forward measures, up to some 
overall parametric or functional freedom. This freedom is used to calibrate the
model to the prices of derivatives, typically interest rate caps and swaptions.
The general line of attack outlined above has in one form or another
been widely implemented by financial institutions, and has been in use for
more than two decades.

The volatility approach does nevertheless suffer from various defects,
conceptual and practical, and it is reasonable to ask if one can do better. We
shall not attempt a detailed critique here, but the following points can be made.
One problem with the volatility approach is that it is difficult to impose a
transparent condition on the discount bond volatility structure that ensures
interest rate positivity (see, e.g., Brody \& Hughston 2002). There is no clear
economic motivation for choosing one volatility structure over another, and
the fact that the volatility is modelled in the risk-neutral measure (or some
other ``unnatural" measure) further removes the model from economic
reality. In this respect the elimination of the market price of risk is ultimately
a shortcoming rather than a virtue. Originally it was thought that (following
the triumph of the Black-Scholes formula) the lack of any need to model the
actual returns of bonds was a ``good thing". The argument was that market
volatilities could be inferred from option prices, whereas market returns
could not, and therefore it would save a lot of trouble if one could avoid
having to model the latter. But those were the days in which ``derivative risk
management" was mostly about pricing and hedging. The industry is wiser
now, and there is a general perception to the effect that ``buy side" concerns,
which have always been somewhat less well-developed mathematically,
involving the aforementioned  issues of portfolio allocation, scenario
analysis, and forecasting, are just as relevant to the ``sell side" as they are
to their colleagues on the other side of the Chinese wall. This argues for
the reinstatement of $\mathbb P$ and the abolition of $\mathbb Q$.

\section{Pricing kernel approach}

\noindent An alternative to the volatility approach is to base the theory on
pricing kernels. The pricing kernel method allows for interest rate positivity,
and it generalises readily to models not based on Brownian motion. The
connection with economic thinking is clearer, and the extension to other
asset classes (such as foreign exchange or inflation-linked products) is
cleaner. The method is rooted in $\mathbb P$, so that although the use of
other measures arises naturally enough  in the course of various specific
calculations there is no temptation to model ``in the risk neutral measure"
from the outset, a short cut that has often been taken in industry
implementations in the past, but is in the final analysis limiting. It follows
also that $\{\lambda_t\}$ is present all along in the pricing kernel approach
as part of the modelling framework, and is not swept underneath a
$\mathbb Q$-rug.

The idea is as follows. We assume the absence of arbitrage opportunities,
but not market completeness. The filtration $\{\mathcal{F}_t\}$, which we
take to satisfy the ``usual conditions", need not be Brownian, so jumps can
be accommodated. Asset price processes have the c\`adl\`ag property. We
assume the existence of an established pricing kernel
$\{\pi_t\}_{t\geq0}$ satisfying $\pi_t >0$ almost surely for $t\geq0$, and
such that for any asset with price process $\{S_t\}_{t\geq0}$ and cumulative
dividend process $\{\Delta_t\}_{t\geq0}$, the associated ``deflated" or
``discounted, risk-adjusted" price process $\{\bar{S}_t\}_{t\geq0}$ defined by
\begin{equation}
\label{axiom}
\bar{S}_t = \pi_t S_t + \int_0^t \pi_s \,\rd \Delta_s
\end{equation}
is a $\mathbb{P}$-martingale. Thus if $\{S_t\}_{t\geq0}$ represents the
price of an asset that pays no dividend, then $\{\pi_t S_t\}_{t\geq0}$ is a
martingale.
If an asset delivers a single random cash flow $H_T$ at
$T$, and derives its value from that cash flow, then its value at $t<T$ is
\begin{eqnarray}\label{pricing formula}
S_{t} = \frac{1}{\pi_t}\, {\mathbb E}[\pi_T H_T\vert\mathcal{F}_t] .
\end{eqnarray}
In the case of a discount bond, which generates a cash flow
of unity at $T$, we have
\begin{eqnarray}\label{bond price}
P_{tT} =  \frac{1}{\pi_t}\, {\mathbb E}[\pi_T \vert\mathcal{F}_t]
\end{eqnarray}
for $t<T$. It follows that we can use the pricing kernel as a basis for interest
rate modelling. In particular, if we model $\{\pi_t\}$ parametrically, then we can
generate families of bond price processes, and use the resulting freedom to
calibrate the model to the prices of select market instruments, as in the volatility
approach.

According to Ushbayev (2011), the notion of a ``pricing kernel" dates back to 
the 1970s, and is used  for example  in Ross (1978), who has apparently
modified the term ``market kernel" used by Garman (1976). Authors have employed
a variety of terms for essentially the same concept. Economists often speak of 
the ``marginal rate of substitution".  The term ``state price density"  appears in 
Dothan \& Williams (1978). One finds the term ``stochastic discount factor" in 
Cox \& Martin (1983), whereas the term ``state price deflator" is used by Duffie 
(1992). 

The idea of using the pricing kernel as a basis for interest rate modelling appears
rather explicitly in Constantinides (1992). The following brief excerpt from this
reference is indicative of the point of view proposed therein (we have changed his
notation slightly to conform with ours): ``We assert the existence of a positive
state-price density or pricing kernel $\{\pi_t\}$ such that the nominal price at time
$t$ of a claim to a nominal payoff $H_T$ at some future date $T$ is given by
[our equation] (\ref{bond price}), where $ {\mathbb E}[ - \vert\mathcal{F}_t]$
denotes the expectation conditional on the information at time $t$. The $\dots$
approach taken here is to explore directly the time-series process of $\{\pi_t\}$,
which yields plausible implications about the term structure of interest rates. I
stress that in this $\ldots$ approach it is unnecessary to assume a representative
consumer economy in which the consumer has von Neumann-Morgenstern
preferences."

There are several different but more or less equivalent ways of representing
the structure of the pricing kernel. Perhaps the most straightforward is to regard
the short rate and the market price of risk as being the primitive data, and
write the pricing kernel in the form
\begin{eqnarray} \label{pricing kernel formula}
\pi_t =  \exp \left( - \int_0^t r_s \, {\rd} s- \int_0^t \lambda_s
\, {\rd} W_s - \half \int_0^t \lambda_s^2 \, {\rd} s \right).
\end{eqnarray}
This line of attack works particularly well in the case of the ``classical" short rate
models, such as those of Vasicek (1977) and Cox {\em et al.}~(1985), where one 
typically starts with an ansatz for the interest rate and the market price of risk. 
Then one can deduce the bond prices by use of (\ref{bond price}), and the 
prices of derivatives based on bond prices (i.e.~interest rate derivatives) by use 
of (\ref{pricing formula}). The difficulty of this approach is that from an economic 
perspective it is unnatural to model the interest rate and market 
risk aversion processes ``separately". In economic analysis, these typically go 
hand in hand. Furthermore, the initial term structure is buried away in the 
specification of the primitive date, and there is no obvious prescription for 
calibrating the model, so one has to proceed on a case by case basis. Indeed, 
this is just what practitioners did before the advent of the HJM method.

We are thus led to ask the following question: do we actually lose anything 
by adopting the pricing kernel method, as opposed to the volatility method? 
In a Brownian setting the answer is no. In fact, the pricing kernel itself can be 
expressed in terms of the volatility $\{\Omega_{tT}\}$, the market price of risk 
$\{\lambda_t\}$, and the initial bond prices $\{P_{0t}\}$, as follows:
\begin{eqnarray}  \label{pricing kernel formula 2}
\pi_t = P_{0t} \exp \left( \int_0^t (\Omega_{st}-\lambda_s)
\, {\rd} W_s - \half \int_0^t (\Omega_{st}-\lambda_s)^2 \, {\rd} s \right).
\end{eqnarray}
This can be proved if, making use of (\ref{money market account}), one inserts 
(\ref{eq:4}) into (\ref{pricing kernel formula}) and simplifies the result (see Jin 
\& Glasserman 2001 and Tsujimoto 2010). Thus, any Brownian ``volatility 
model" can be converted into a ``pricing kernel model" and vice-versa. More 
precisely, we can regard $\{\Omega_{tT}\}$, $\{\lambda_t\}$, and $\{P_{0t}\}$ 
as being specified up to some overall parametric freedom, thus inducing a 
corresponding parametrisation of the pricing kernel, which can then be 
calibrated to market data and/or market forecasts by various schemes. 

The upshot of this is that the various ``approaches" to modelling interest rates 
amount to different ways of representing the pricing kernel, such as the formulae
given by (\ref{pricing kernel formula}) and (\ref{pricing kernel formula 2}), 
which in turn suggest various distinct way of parameterising the resulting 
models. We shall, in what follows, consider yet another representation of the 
pricing kernel, namely that associated with the so-called Flesaker-Hughston 
(FH) models. In this representation (Flesaker \& Hughston 1996, 1997, 
1998) the pricing kernel takes the form
\begin{eqnarray}\label{pk}
\pi_t = \int_t^\infty (-\partial_s P_{0s}) M_{ts} \, {\rd} s,
\end{eqnarray}
where $\{M_{ts}\}_{s\geq t\geq0}$ is a family of positive unit-initialised martingales.
Thus, we require that $M_{0s}=1$ for $s\geq0$, that $M_{ts}>0$ for 
$0\leq t\leq s<\infty$, and that
$\mathbb{E}[M_{us}\vert\mathcal{F}_t]=M_{ts}$
for $0\leq t\leq u\leq s<\infty$. It follows from the assumed asymptotic property of
the initial term structure, without further restriction on the martingale family, that
the right side of (\ref{pk}) is finite almost surely. In particular, by use of
the Fubini theorem, the martingale property, and the fundamental theorem of calculus, 
one finds that $ \mathbb{E}[\pi_t]=P_{0t} $.
As a consequence of equation (\ref{bond price}) we deduce that the
discount bond system takes the rational form
\begin{eqnarray}\label{eq:7}
P_{tT} = \frac{\int_T^\infty (-\partial_s P_{0s}) M_{ts} \, {\rd} s}
{\int_t^\infty (-\partial_s P_{0s}) M_{ts} \, {\rd} s}.
\end{eqnarray}
To model the interest rate system we thus need to specify the initial term structure
$\{P_{0t}\}$ together with a family of positive martingales. 
The FH model originated as an attempt to characterise the 
complete family of HJM-type models with positive interest rates and valid over 
an arbitrary time horizon. The work was presented at the Cornell/Queen's 
University Derivative Securities conference organised by R.~Jarrow and 
S.~Turnbull in April 1995 (cf.~Hughston 2003). A generalised version of the 
model was presented by Rutkowski (1997), and it was made clear in this later 
work that the model was not tied to the use of a Brownian filtration. In the words 
of Rutkowski (1997): ``From a theoretical viewpoint, the basic input of the 
generalised Flesaker-Hughston model is a strictly positive supermartingale 
\dots". The relation of the FH models to the HJM theory and other approaches 
has been discussed in detail by a number of authors, including 
Jin \& Glasserman (2001), Cairns (2004), Hunt \& Kennedy (2004), Musiela 
\& Rutkowski (2005), and Bjork (2009).  Other representations 
of the pricing kernel method that offer interesting insights include the use of 
potentials (Rogers 1997), and Wiener chaos (Brody \& Hughston 
2004, Hughston \& Rafailidis 2005); these will not be considered here. 
We return to the FH representation in the context of L\'evy models.

\section{ L\'evy models for interest rates}

\noindent One important deficiency of the volatility approach to interest rate 
modelling is that in its more successful implementations it has been so deeply 
intertwined with Brownian motion based modelling techniques that little by way of
consensus has emerged either in the industry or among academics on how 
best to incorporate jumps into the scheme. This being the case, it will not be 
amiss here to attempt to make some progress on the matter. 

The theory of L\'evy models for asset pricing has an extensive literature. We 
refer the reader, for example, to Madan \& Senata (1990), Madan \& Milne 
(1991), Heston (1993), Gerber \& Shiu (1994), Eberlein \& Keller (1995), Eberlein 
\& Jacod (1997), Chan (1999),  Kallsen \& Shiryaev (2002), Fujiwara \& 
Miyahara (2003), Schoutens (2003), Cont \& Tankov (2004), Esche \& 
Schweizer (2005), Hubalek \& Sgarra (2006), 
and references cited therein. By a L\'evy process on a probability space 
$(\Omega,{\mathcal F},{\mathbb P})$ we mean a process $\{X_t\}$ such that 
$X_0=0$, $X_t-X_s$ is independent of ${\mathcal F}_s$ for $t\geq s$, and 
${\mathbb P}(X_t-X_s\leq y) = {\mathbb P}(X_{t+h}-X_{s+h}\leq y)$ for $h>0$. Here 
$\{{\mathcal F}_t\}$ denotes the augmented filtration generated by $\{X_t\}$. 
For $\{X_t\}$ to give rise to a L\'evy model for asset prices, we require that it 
should possess exponential moments; that is,  
\begin{eqnarray}
{\mathbb E}[\re^{\alpha X_t}] < \infty  
\end{eqnarray}
for $t\geq 0$, for $\alpha$ in some connected real interval $A$ containing 
the origin. The stationary and independent increments property implies
that there exists a L\'evy exponent $\psi(\alpha)$ such that 
\begin{eqnarray}
{\mathbb E}[\re^{\alpha X_t}] = \re^{t\psi(\alpha)}
\label{cumulant}
\end{eqnarray}
for  $\alpha\in A$. Then the process
defined by
\begin{eqnarray}
M_t = \re^{\alpha X_t-t\psi(\alpha)}   
\label {martingale}
\end{eqnarray}
is a martingale, called the associated geometric L\'evy 
martingale (or Esscher martingale), with parameter 
$\alpha$. More generally, let $\{\alpha_t\}$ be
an $\{{\mathcal F}_t\}$--predictable process, chosen in such a way 
that $\alpha_t\in A$ for $t\geq0$, and such that the local martingale defined by
\begin{eqnarray}
M_t = \exp \left({\int_0^t \alpha_s {\rd} X_s-\int_0^t\psi(\alpha_s){\rd} s}\right)
\end{eqnarray}
is a martingale. If a predictable process $\{\alpha_t\}$ satisfies these 
conditions then we say it is admissible. Then we are led to consider an 
asset pricing model of the 
following form. Let the exogenously specified  short rate $\{r_t\}$ be 
$\{{\mathcal F}_t\}$--adapted, and be such that the unit-initialised money 
market account defined as in
(\ref{money market account})
is finite almost surely for $t>0$. Let the $\{{\mathcal F}_t\}$-adapted risk 
aversion and volatility processes $\{\lambda_t\}$ and $\{\sigma_t\}$ be positive, 
and be such that $\{-\lambda_t\}$, $\{\sigma_t\}$, and 
$\{\sigma_t -\lambda_t\}$ are admissible in the sense noted above. The 
pricing kernel is taken to be given by 
\begin{eqnarray}
\pi_t = \exp \left({-\int_0^t r_s\, {\rd} s -\int_0^t \lambda_s \, {\rd} X_s
-\int_0^t\psi(-\lambda_s)\,{\rd} s}\right),
 \label{general pricing kernel}
\end{eqnarray}
and the associated expression for the price of a typical non-dividend-paying 
asset is 
\begin{eqnarray}
S_t = S_0 \exp \left({\int_0^t r_s \,{\rd} s + \int_0^t R(\lambda_s,\sigma_s)\, 
{\rd} s + \int_0^t \sigma_s \,{\rd} X_s-\int_0^t\psi(\sigma_s)\,{\rd} s}\right),
\label{general asset}
\end{eqnarray}
where 
\begin{eqnarray}
R(\lambda,\sigma) = \psi(\sigma) + \psi(-\lambda) - \psi(\sigma-\lambda). 
\label{R}
\end{eqnarray}
is the excess rate of return function associated with the given L\'evy 
exponent. It is a remarkable property of the function $R(\lambda,\sigma)$, 
arising as a consequence of the convexity of the L\'evy exponent, that if 
the volatility and the risk aversion are positive, then the excess rate of 
return is positive, and is monotonically increasing in its arguments
(Brody {\em et al.}~2011). In general the excess rate of return is 
nonlinear, and hence in a general L\'evy setting the process 
$\{\lambda_t\}$ no longer admits the interpretation of being the 
market price of risk, but instead should be viewed as a measure of 
investor risk aversion.

The theory of interest rates can be developed in just this spirit in a way that  
generalises the HJM framework to the L\'evy category.  In particular, we 
are able to give a consistent treatment of the risk premium associated with 
interest rate products in such a way that the risk premium is positive for 
bonds. Interest rate models admitting jumps
have been pursued by a number of authors, including,  {\em inter alia},
Shirakawa (1991), Jarrow \& Madan (1995), Bj\"ork \textit{et al}.  (1997), 
Bj\"ork \textit{et al}. (1997), Eberlien \& Raible (1999), Raible (2000),  
Eberlein \textit{et al}. (2005), Eberlein \& Kluge  (2006a,b, 2007), and 
Filipovi\'c \textit{et al}. (2010), to mention a few. Our approach is novel 
inasmuch as we introduce a pricing kernel at the outset, rather than 
attempting to model the interest rate system through a set of dynamical 
equations. A rather general class of L\'evy interest rate models exhibiting 
the positive excess rate of return property can thus be constructed as 
follows. The idea is to model  the pricing kernel and the associated 
discount bond system. There is no need to introduce a system 
of instantaneous forward rates. Let the pricing kernel be given by 
(\ref{general pricing kernel}), and write $P_{tT}$ for the price at $t$ of a bond 
that matures at $T$ to deliver one unit of currency. For the L\'evy discount 
bond model we  have:
\begin{eqnarray}
P_{tT} = P_{0T} \exp\left( \int_0^t r_s \rd s + \int_0^t R(\lambda_s,\Omega_{sT})
\rd s + \int_0^t \Omega_{sT} \rd X_s - \int_0^t \psi(\Omega_{sT} )\rd s \right).
\end{eqnarray}
We require (a) that $\{\lambda_t\}$ and $\{\Omega_{tT}\}$ are positive, (b) that 
$\{-\lambda_t\}$,   $\{\Omega_{tT}\}$, and $\{\Omega_{tT} -\lambda_t \}$, 
are admissible in the sense indicated above, and (c) that  $\{\Omega_{tT}\}$ 
should vanish as $t$ approaches $T$. 
The maturity 
condition $\lim_{t \to T}P_{tT} =1$ allows one to solve for the money market 
account in terms of $\{\Omega_{tT}\}$ and $\{\lambda_t\}$, as in the Brownian case:
\begin{eqnarray}
B_t = \frac{1}{P_{0t}} \exp\left(  -\int_0^t \Omega_{st} \rd X_s + \int_0^t 
\psi(\Omega_{st}-\lambda_s)\rd s  - \int_0^t \psi(-\lambda_s)\rd s \right) .
\end{eqnarray}
Inserting the expression for the money market account back into the bond price, 
we obtain:
\begin{eqnarray}
P_{tT}=P_{0tT} 
\frac{\exp\left(\int_0^t (\Omega_{sT}-\lambda_s)\rd X_s - \int_0^t \psi(\Omega_{sT}
-\lambda_s)\rd s\right)}
{\exp\left(\int_0^t (\Omega_{st}-\lambda_s)\rd X_s - \int_0^t \psi(\Omega_{st}
-\lambda_s)\rd s\right)},
\end{eqnarray}
where $P_{0tT}=P_{0T}/P_{0t}$. If we substitute the expression for the 
money market account back into equation (\ref{general pricing kernel}), and 
simplify the result, we deduce the following:

\begin{prop}\label{prop:0}
The pricing kernel in a L\'evy interest rate model can be expressed in terms of 
the initial term structure $\{P_{0t}\}$, the bond volatility $\{\Omega_{tT}\}$, and 
the risk aversion $\{\lambda_t\}$ as:
\begin{eqnarray}
\pi_t = P_{0t} \exp\left(\int_0^t (\Omega_{st}-\lambda_s)\rd X_s - \int_0^t 
\psi(\Omega_{st}-\lambda_s)\rd s\right) .
\end{eqnarray}
\end{prop}

This result offers one a rather general method for modelling an arbitrage-free 
interest rate system in the physical measure when there are jumps. 
In particular, if we model  the volatility structure and the risk aversion
exogenously, and specify the initial term structure, then we determine 
the pricing kernel, the discount bond system, and the money market account. 
In practice, as in the Brownian case, the volatility structure and the risk aversion process can be
modelled parametrically, up to some undetermined functional degrees of 
freedom, to be fixed by calibration to the prices of market instruments 
at time zero. Indeed, one could treat the L\'evy exponent itself as part of the 
``functional freedom" of the model. 

One can also consider an extension of the FH models to the L\'evy category
as a basis for representing the pricing kernel. 
We shall construct the required martingale families from the
Esscher martingales associated with various L\'evy processes. Thus we 
fix a probability space $(\Omega,\mathcal{F},\mathbb{P})$ and a L\'evy 
process
$\{X_t\}_{t\geq0}$ admitting exponential moments as before, and for a suitable deterministic function
$ \{ \phi_t \}_{ t\geq 0} $ we define a martingale family
$\{M_{ts}\}$ by setting
\begin{eqnarray}\label{m}
M_{ts}=\frac{{\rm e}^{\phi_s X_t}}{\mathbb{E}[{\rm e}^{\phi_s X_t}]}.
\end{eqnarray}
We note that  $M_{ts}>0$ and $M_{0s}=1$. We require 
that  $\phi_s\in A$ for all $t \geq 0 $. By making various choices for $\{X_t\}$ we are thus able to
generate a variety of interest rate models, each with some functional freedom
given by the choice of $\{\phi_t \}$. 
In the general setting, both $ \{\phi_s \}$ and $\{X_t\}$
are vectorial, and the components of $\{X_t\}$ are independent processes. For the present
we look at one-dimensional models. In Section \ref{sgbm} we consider first the case of a geometric Brownian motion 
(GBM) family. In this instance there are of course no jumps, but the calculations involved are indicative 
of the more general case, and are of some interest in their own right. 
Expressions are derived for discount bonds, the short 
rate, the  bond volatility, and the risk aversion. In Section \ref{srp} 
we establish necessary and sufficient conditions on $\{\phi_t\}$ to ensure the 
positivity of the risk premium in the case of the GBM family, and in Section 
\ref{sop} we solve the associated option pricing problem. In Sections 
\ref{sjd}, \ref{sg}, and \ref{svg} we proceed to construct models based on the 
Esscher martingale families associated with jump-diffusion 
processes, gamma processes, and variance-gamma processes, and to examine their properties.

\section{GBM family}\label{sgbm}

\begin{figure}[t!]
\centering
\begin{center}
\includegraphics[width=8cm,height=10cm,keepaspectratio]{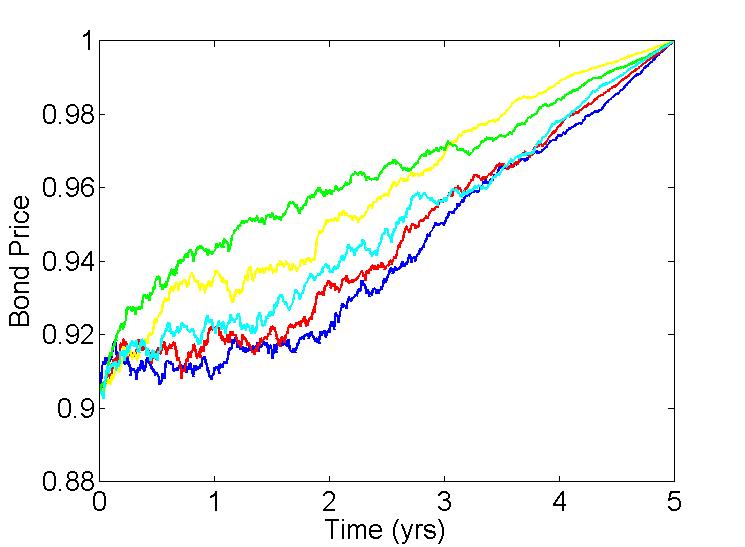}
\includegraphics[width=8cm,height=10cm,keepaspectratio]{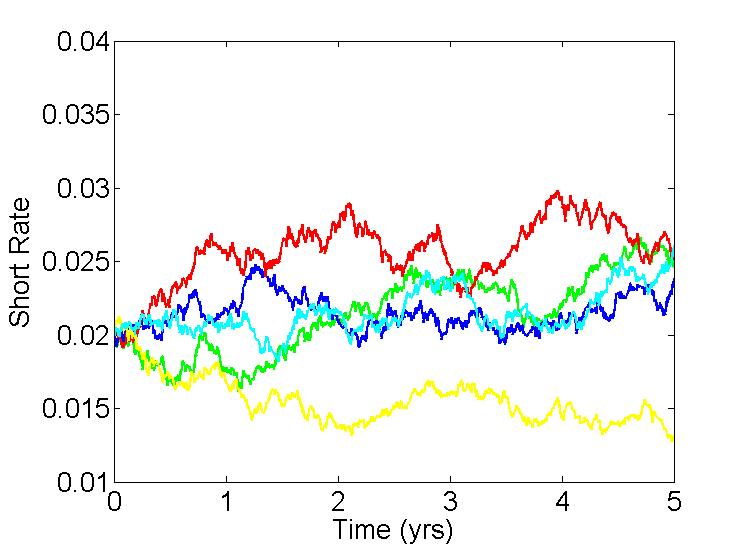}
\caption{ \textit{GBM family}. 
Simulations of the bond price process (\ref{gbbp}) and the short rate process 
(\ref{gbsr}) in a rational term-structure model with a parametric martingale family 
based on a geometric Brownian motion. The bond maturity is five years, the initial 
term structure is assumed to be flat so that $P_{0t}={\rm e}^{-0.02t}$, and we set 
$\phi_s =0.3{\rm e}^{-0.02s}$.}\label{figbm}
\end{center}
\end{figure}

\noindent Writing $\{W_t\}_{t\geq0}$ for a standard Brownian 
motion, we obtain a geometric 
L\'evy martingale family of the form
\begin{equation}
M_{ts}=\exp\left(\phi_s W_t-\half \phi_s ^2t\right)
\label{eq:30}
\end{equation}
for $0\leq t\leq s$. 
This leads to the following system of discount bond prices:
\begin{eqnarray}\label{gbbp}
P_{tT}=\frac{\int_T^\infty\rho_s \exp(\phi_s W_t-\frac{1}{2}\phi_s ^2t)\,
{\rd} s} {\int_t^\infty\rho_s \exp(\phi_s W_t-\frac{1}{2}\phi_s ^2t)\,{\rd} s}
\end{eqnarray}
for $0\leq t\leq T<\infty$ (cf.~Brody \& Friedman 2009).
The function $\{\rho_t \}_{t\geq0}$ denotes the initial
``term-structure density" (Brody \& Hughston 2001), and is given by
$
\rho_t  = -\partial_t P_{0t}.
$
We make note of the fact that if the initial interest rates are positive and if
$\lim_{t\to\infty}P_{0t}=0$, then $\{\rho_t \}$ fulfils the requirements of a density
function: namely, that $\rho_s >0$ for $s\geq0$, and that
\begin{equation}
\int_0^\infty\rho_s \, {\rd} s=1.
\end{equation}
By use of the relation 
$r_t= - \lim_{s\to t } \partial_t P_{st}$
it follows that the short rate is of the form
\begin{eqnarray}\label{gbsr}
r_t=\frac{\rho_t \exp(\phi_t W_t-\frac{1}{2}\phi_t ^2t)} {\int_t^\infty\rho_s 
\exp(\phi_s W_t-\frac{1}{2}\phi_s ^2t)\,{\rd} s}.
\end{eqnarray}
Simulations of the bond price (\ref{gbbp}) and the short rate (\ref{gbsr}) are
presented in Figure \ref{figbm}. We note that $\{r_t\}$ need not in general be 
bounded from above. Some rational term-structure models, such as the rational lognormal model (Flesaker \& Hughston 1996), and the separable 
second-chaos models (Hughston \& Rafailidis 2005)  have
bounded interest rates. See Cairns (2004) for further discussion on this point.
By an application of Ito's lemma to equation (\ref{gbbp}), we deduce that the
dynamical equation of the bond price system is given by
\begin{eqnarray}\label{pdynamics}
\frac{{\rd} P_{tT}}{P_{tT}}=(r_t-\Phi_{tt}(\Phi_{tT}-\Phi_{tt}))\,{\rd} t+
(\Phi_{tT}-\Phi_{tt})\,{\rd} W_t,
\end{eqnarray}
where
\begin{eqnarray}\label{pbig}
\Phi_{tT}=\frac{\int_T^\infty\phi_s \rho_s \exp\left(\phi_s W_t-\frac{1}{2}\phi_s ^2t
\right){\rd} s} {\int_T^\infty\rho_s \exp\left(\phi_s W_t-\frac{1}{2}\phi_s ^2t\right)
{\rd} s},
\end{eqnarray}
and $\Phi_{tt}=\lim_{s \to t} \Phi_{st}$.
We note that the bond volatility is of the form
$
\Omega_{tT}=\Phi_{tT}-\Phi_{tt},
$
and that the market price of risk is given by
$
\lambda_t=-\Phi_{tt}.
$
The bond volatility and the market price of risk together give us the risk
premium associated with an investment in a discount bond:
\begin{eqnarray}\label{riskprem}
\lambda_t  \Omega_{tT}=\Phi_{tt}\left(\Phi_{tt}-\Phi_{tT}\right).
\end{eqnarray}
Since $\{\Omega_{tT}\}$ and $\{\lambda_t\}$ are determined by 
$\{\Phi_{tT}\}$, which in turn is determined by $\{\rho_t\}$ and $\{\phi_t\}$ 
via (\ref{pbig}), we  conclude that the model is fully characterised by the 
specification of the initial term structure $\{\rho_t\}$ and the martingale 
volatility structure $\{\phi_t\}$. It is worth remarking that even if the 
volatility $\{\phi_t\}$ of the martingale family (\ref{eq:30}) is deterministic, 
both the bond price and the short rate exhibit nontrivial 
stochastic dynamics.

\section{Positivity of risk premium in the GBM family}\label{srp}

\noindent In the case of a volatility-based representation for the pricing kernel we were able to 
identity the condition for the positivity of the excess rate of return above the 
short rate. That is, we require that $R(\lambda_t,\Omega_{tT})>0$, where 
$R(\lambda,\sigma)$ is defined by (\ref{R}), and this can be satisfied if we specify exogenously that
the risk aversion and the volatility are positive.  In the case of an FH model, it
 is not immediately apparent from  expression (\ref{riskprem}) 
for the excess rate of return that the required conditions are satisfied. 
Since $\{\rho_t\}$ is fixed by the initial term 
structure, the relevant condition for the positivity of the excess rate of return must 
be imposed on the choice of the functional model parameter $\{\phi_t\}$. 
We shall establish necessary and sufficient conditions on 
$\{\phi_t\}$ such that $\lambda_t\Omega_{tT}>0$. 

\begin{prop}\label{prop:1}
In an FH model based on a GBM family with functional model parameter $\{\phi_t \}$, a sufficient condition to 
ensure that the risk premium is positive is that either $\{\phi_t \}$ should be  positive and 
decreasing, or that $\{\phi_t \}$ should be  negative and increasing.
\end{prop}
\noindent {\bf Proof}. 
Suppose that $\phi_s $ is positive for all $s\geq0$. Then it follows from
(\ref{pbig}) that $\Phi_{tt}$ is positive. Differentiating (\ref{pbig}) with respect
to $T$ we obtain
\begin{eqnarray}\label{pdifT}
\partial_T\Phi_{tT}=f_{tT}\left(\Phi_{tT}-\phi_T \right),
\label{eq:31}
\end{eqnarray}
where
\begin{eqnarray}
f_{tT}=\frac{\rho_T \exp\left(\phi_T W_t-\frac{1}{2}\phi_T^{\,2}t\right)}
{\int_T^\infty\rho_s \exp\left(\phi_s W_t-\frac{1}{2}\phi_s ^{\,2}t\right){\rd} s}
\end{eqnarray}
is the instantaneous forward rate defined as usual by
$
f_{tT}=-\partial_T\ln P_{tT},
$
which is positive in any FH model. Next we observe that
(\ref{pbig}) can be written in the form
\begin{eqnarray}\label{pmu}
\Phi_{tT}=\int_T^\infty\phi_s \, \mu_{tT}(s){\rd} s,
\end{eqnarray}
where
\begin{eqnarray}
\mu_{tT}(s)=\frac{\rho_s \exp\left(\phi_s  W_t-\frac{1}{2}\phi_s ^2t\right)}
{\int_T^\infty\rho_s \exp\left(\phi_s W_t-\frac{1}{2}\phi_s ^2t\right){\rd} s}.
\label{eq:35}
\end{eqnarray}
Note that $\mu_{tT}(s)$ is positive and that
\begin{eqnarray}
\int_T^\infty\mu_{tT}(s){\rd} s=1.
\end{eqnarray}
Thus, according to (\ref{pmu}), $\Phi_{tT}$ is a weighted average of the values
of $\phi_s $ for $s$ greater than or equal to $T$. It follows that if $\phi_s $ is
decreasing as a function of $s$, then $\Phi_{tT}<\phi_T$, for $0<t<T$. This in
turn implies, by use of (\ref{pdifT}), that $\partial_T\Phi_{tT}<0$, and hence by
(\ref{riskprem}) that the risk premium is positive. A similar argument shows that
if $\phi_s $ is negative and increasing for $s\geq0$, then the risk premium is 
positive. \hfill$\Box$
\vspace{0.2cm}

\begin{prop}\label{prop:2}
In an FH model based on a GBM family with functional model parameter $\{\phi_t \}$,
assume that for any admissible initial term-structure density $\{\rho_t\}$ the risk premium  
is positive. Then $\{\phi_t\}$ must be 
either positive and decreasing, or  negative and
increasing.
\end{prop}

\noindent {\bf Proof}.  If $\Phi_{tt}(\Phi_{tt}-\Phi_{tT})>0$, then either (i) 
$\Phi_{tt}>0$ and $\Phi_{tT}-\Phi_{tt}<0 $ holds; or (ii) $\Phi_{tt}<0$ and 
$\Phi_{tT}-\Phi_{tt}>0 $ holds.  Assume that (ii) holds.
Let us define a random probability measure $\mu_t(s)$ by setting 
$\mu_t(s)=\mu_{tt}(s)$ where $\mu_{tT}(s)$ is defined in (\ref{eq:35}). Note that $\{\mu_t(s)\}_{s\geq t}$ has the property that $\mu_0(s)=\rho_s$. 
We can then express the condition $\Phi_{tt}<0$ in the form 
\begin{eqnarray}
\int_t^\infty \phi_s \mu_t(s) \rd s < 0,
\end{eqnarray}
which must hold for all $0\leq t<T\leq\infty$. Specifically, at $t=0$ we have 
\begin{eqnarray}
\int_0^\infty \phi_s \rho_s \rd s < 0.
\end{eqnarray}
Since this has to hold for an arbitrary density $\rho_s$, it has to hold, in particular, 
for the limiting case of a delta function $\rho_s=\delta(s-t)$ for any $t>0$. It follows that $\phi_s$ 
must be negative for all $s\geq 0$. Next we observe that since $\Phi_{tT}-\Phi_{tt}>0$  holds by assumption, the fact that this inequality must hold, in particular, for $T$ 
slightly greater than $t$ implies that 
\begin{eqnarray}
\left. \frac{\partial \Phi_{tT}}{\partial T} \right|_{T=t} > 0 
\end{eqnarray}
for all $t\geq 0$. By (\ref{eq:31}) and the fact that the instantaneous forward rates 
are positive, we deduce that $\Phi_{tt}>\phi_t$ for all $t\geq 0$. Therefore, we
obtain 
\begin{eqnarray}
\int_t^\infty \phi_s \mu_t(s) \rd s > \phi_t 
\end{eqnarray}
for all $t\geq 0$. Now letting $\rho_s = \delta(s-u)$ for any $u>t$, we deduce that 
$\phi_u > \phi_t$. Similar arguments can be used in the case 
where (i) $\Phi_{tt}>0$ and $\Phi_{tT}-\Phi_{tt}<0$ holds. \hfill$\Box$

\section{Option Pricing in the GBM case}\label{sop}

\noindent We consider the option pricing problem in the case of the geometric 
Brownian motion family. First we discuss the problem of option pricing in a
general L\'evy model, and then we specialise to the Brownian case. The price at time 
$0$ of a European call option expiring at time $t$, with strike price $K$, on a 
discount bond maturing at time $T$, is given by
\begin{eqnarray}\label{o}
C_{0t}=\mathbb{E}[\pi_t(P_{tT}-K)^+].
\end{eqnarray}
We assume that $0<K<1$. By use of (\ref{bond price}) it follows that
\begin{eqnarray}
C_{0t}=\mathbb{E}\left[\left(\mathbb{E}_t[\pi_T]-K\pi_t\right)^+\right],
\end{eqnarray}
and hence by (\ref{eq:7}) we have
\begin{eqnarray}\label{optpr}
C_{0t}=\mathbb{E}\left[\left(\int_T^\infty\rho_s M_{ts}{\rd} s-K\int_t^\infty\rho_s 
M_{ts}{\rd} s\right)^+\right].
\end{eqnarray}
Recall that the martingale family appearing in (\ref{optpr}) takes the form 
\begin{eqnarray}
M_{ts}=\exp\left(\phi_s X_t-\psi(\phi_s)t\right).
\end{eqnarray}
For an option price in the general setting we thus obtain
\begin{eqnarray}\label{genop}
C_{0t}=\mathbb{E}\left[\left(\int_T^\infty\rho_s {\re}^{\phi_s X_t-\psi(\phi_s)}
{\rd} s-K\int_t^\infty\rho_s {\re}^{\phi_s X_t-\psi(\phi_s)t}{\rd} s\right)^+\right].
\end{eqnarray}
We observe that if $\{\phi_s \}$ is decreasing in $s$, then the function $P(t,T,\xi)$
defined by
\begin{eqnarray}
P(t,T,\xi)=\frac{\int^\infty_T\rho_s \exp\left(\phi_s \xi-\psi(\phi_s)t\right){\rd} s}
{\int^\infty_t\rho_s \exp\left(\phi_s \xi-\psi(\phi_s)t\right){\rd} s}
\end{eqnarray}
is decreasing in the variable $\xi$. The argument is as follows.
A short calculation shows that
\begin{eqnarray}\label{ln}
\frac{\partial\ln P(t,T,\xi)}{\partial\xi}=\Phi_{tT}(\xi)-\Phi_{tt}(\xi),
\end{eqnarray}
where the function $\Phi_{tT}(\xi)$ is defined by
\begin{eqnarray}
\Phi_{tT}(\xi)=\frac{\int^\infty_T\phi_s \rho_s \exp\left(\phi_s \xi-\psi(\phi_s)
t\right){\rd} s} {\int^\infty_T\rho_s \exp\left(\phi_s \xi-\psi(\phi_s)t\right){\rd} s}.
\end{eqnarray}
We observe that if $\{\phi_s \}$ is decreasing then
\begin{eqnarray}
\frac{\partial\Phi_{tT}(\xi)}{\partial T}<0.
\end{eqnarray}
This follows from the fact that
\begin{eqnarray}\label{part}
\frac{\partial\Phi_{tT}(\xi)}{\partial T}=f_{tT}(\xi)\left(\Phi_{tT}(\xi)-\phi_T \right),
\end{eqnarray}
where
\begin{eqnarray}
f_{tT}(\xi)=\frac{\rho_T \exp\left(\phi_T \xi-\psi(\phi_T)t\right)}
{\int_T^\infty\rho_s \exp\left(\phi_s \xi-\psi(\phi_s)t\right){\rd} s}.
\end{eqnarray}
We note that $\Phi_{tT}(\xi)$ is for each value of $\xi$ a weighted average of
$\phi_s $ for $s\geq T$. Thus, if $\phi_s $ is decreasing, then the right-hand
side of equation (\ref{part}) is negative. If the right-hand side of (\ref{part}) is
positive ({\em resp}.~negative) then (\ref{ln}) is positive ({\em resp}.~negative).
It follows  that if $\phi_s $ is decreasing in $s$, then $P(t,T,\xi)$ is decreasing
in $\xi$, as claimed. A similar argument shows that if $\phi_s $ is increasing
in $s$ for all $s\geq0$ then $P(t,T,\xi)$ is increasing in $\xi$.

Let us assume now that $P(t,T,\xi)$ is monotonic in $\xi$, and write $P^+(t,T)$
and $P^-(t,T)$ for the upper and lower extremal values of $P(t,T,\xi)$ as $\xi$
varies. Then for any $K$ in the range $[P^-(t,T),P^+(t,T)]$ we can find a number
$\xi^*$ such that
\begin{eqnarray}
P(t,T,\xi^*)=K.
\end{eqnarray}
This enables us to truncate the expectation in (\ref{genop}) at the point where the
maximum function becomes nonpositive. The price of an option in the general 
L\'evy case then takes the form
\begin{eqnarray}\label{eq:OptionPrice}
C_{0t}=\int_T^\infty\rho_s m_{ts}  {\rd} s-K\int_t^\infty\rho_s m_{ts}  {\rd} s,
\end{eqnarray}
where
\begin{eqnarray}
m_{ts}  =\mathbb{E}\left[\Theta\left(\int_T^\infty\rho_s M_{ts}{\rd} s-
K\int_t^\infty\rho_s M_{ts}{\rd} s\right)M_{ts}\right],
\end{eqnarray}
and $\Theta$ is the Heaviside function.
In particular, when the underlying L\'evy martingale is a geometric Brownian 
motion, and $\{\phi_s \}$ is positive and decreasing, the option price simplifies to the
following expression:
\begin{eqnarray}\label{bo}
C_{0t}=\int_T^\infty\rho_s \, N\left(\frac{\xi^*}{\sqrt{t}}-\phi_s \sqrt{t}\right){\rd} s
-K\int_t^\infty\rho_s \,N\left(\frac{\xi^*}{\sqrt{t}}-\phi_s \sqrt{t}\right){\rd} s,
\end{eqnarray}
where $N(x)$ denotes the normal distribution function. We remark 
that when $\{\phi_t\}$ is a strictly monotonic function the resulting discount bond prices and interest 
rates are Markovian. This is because the functions $P(t,T,\xi)$ and $f_{tT}(\xi)$ are 
monotonic functions of $\xi$, and $P_{tT}=P(t,T,X_t)$ and $f_{tT}=f_{tT}(X_t)$.

\section{Geometric jump-diffusion family}\label{sjd}

\begin{figure}[t!]
\centering
\begin{center}
\includegraphics[width=8cm,height=10cm,keepaspectratio]{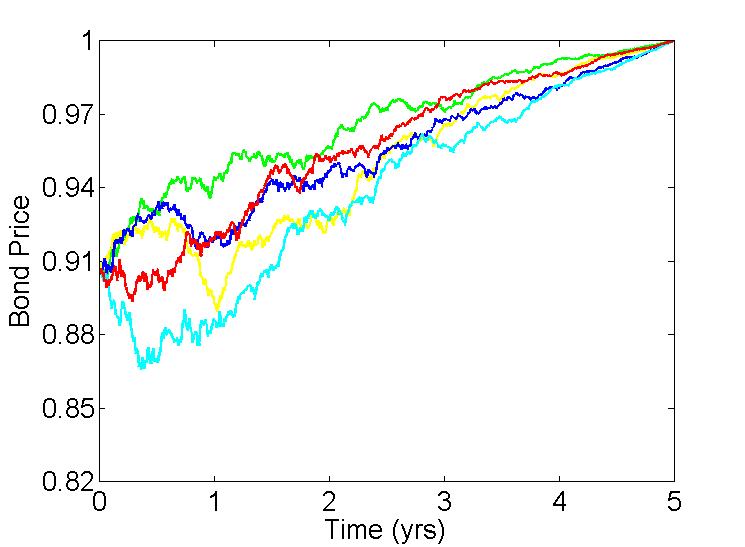}
\includegraphics[width=8cm,height=10cm,keepaspectratio]{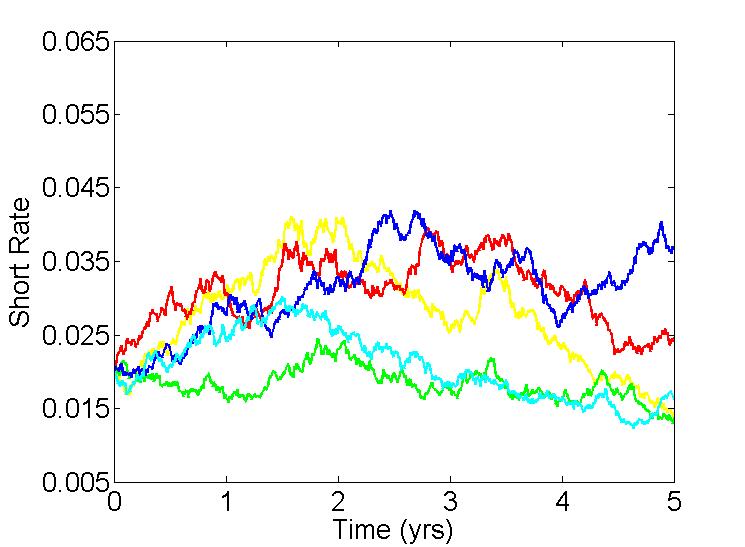}
\caption{\textit{Jump-diffusion family}. 
Simulations of the bond price (\ref{jdbp}) and the short rate (\ref{jdsr}) 
in a rational term-structure model with a parametric martingale family based on 
a geometric jump-diffusion process. The bond maturity is five years, the initial 
term structure is given by $P_{0t}={\rm e}^{-0.02t}$, and we set 
$\phi_s =0.3{\rm e}^{-0.02s}$. The rate parameter of the Poisson process is 
$\lambda=20$. The jump sizes have mean $\mu=0$, and standard deviation 
$\delta=0.09$.}\label{figjd}
\end{center}
\end{figure}

\noindent Merton (1976) extended the Black-Scholes option-pricing theory to 
include equity prices driven by a jump-diffusion process. In term-structure 
modelling it is also desirable to incorporate both jump risk and diffusion 
risk. For simplicity we focus on the case of normally distributed jump sizes.
We introduce a Poisson process $\{N_t\}_{t\geq0}$, with rate parameter 
$\lambda$, to represent the number of jumps occurring by time $t$. The size of 
the $i^{\rm th}$ jump is modelled by a random variable $J_i$. Jump sizes are 
independent and identically distributed random variables, each such that 
$J_i\sim N(\mu,\delta^2)$. If the diffusion component is driven by an 
independent Brownian motion $\{W_t\}$, we obtain an expression for the bond 
price as follows. Writing $\{J_t\}$ for the compound Poisson process defined by
\begin{eqnarray}
J_t=\sum_{i=1}^{N_t}J_i ,
\end{eqnarray}
and introducing a single functional degree of freedom $\phi_s $, by use of
(\ref{m}) we obtain a geometric martingale family of the form
\begin{eqnarray}\label{eq:GeoMartJD}
M_{ts}=\exp\left(\phi_s (W_t+J_t)-\half\phi_s ^2t-\lambda t
({\rm e}^{\phi_s \mu+\frac{1}{2}\phi_s ^2\delta^2}-1 )\right).
\end{eqnarray}
This leads to the following discount bond price:
\begin{equation}\label{jdbp}
P_{tT}=\frac{\int_T^\infty\rho_s \exp \left(\phi_s (W_t+J_t)-\frac{1}{2}
\phi_s ^2t-\lambda t({\rm e}^{\phi_s \mu+\frac{1}{2}\phi_s ^2\delta^2}-1)\right)
{\rd} s} {\int_t^\infty\rho_s \exp \left(\phi_s (W_t+J_t)-\frac{1}{2}\phi_s ^2
t-\lambda t({\rm e}^{\phi_s \mu+\frac{1}{2}\phi_s ^2\delta^2}-1)\right){\rd} s},
\end{equation}
and for the short rate we have
\begin{equation}\label{jdsr}
r_t=\frac{\rho_t \exp\left(\phi_t (W_t+J_t)-\frac{1}{2}\phi_t ^2t-\lambda t
({\rm e}^{\phi_t \mu+\frac{1}{2}\phi_t ^2\delta^2}-1)\right)} {\int_t^\infty\rho_s 
\exp\left(\phi_s (W_t+J_t)-\frac{1}{2}\phi_s ^2t-\lambda t({\rm e}^{\phi_s 
\mu+\frac{1}{2}\phi_s ^2\delta^2}-1)\right){\rd} s}.
\end{equation}
Sample paths of the bond price (\ref{jdbp}) and the short rate (\ref{jdsr}) are
simulated in Figure \ref{figjd}.
In Merton (1976) a key idea used to price options is the notion that
idiosyncratic risk can be modelled with a return equal to the risk-free rate. In
Merton's model it is assumed that the jump risk is purely idiosyncratic and can
be diversified away by holding a suitably broad portfolio. 
 In our model jump risk is being priced. The resulting risk premium is implicit
in the choice of pricing kernel, and is determined by the functional model parameter $\{\phi_t \}$.

To derive an expression for the price of a call option with strike
$K$, we need to evaluate the expectation in (\ref{genop}). By use of the
tower property of conditional expectation we condition on the number of
Poisson jumps $N_t$ to obtain
\begin{eqnarray}
C_{0t}=\mathbb{E}\left[\mathbb{E}\left[\left.\left(\int_T^\infty\rho_s M_{ts}
{\rd} s-K\int_t^\infty\rho_s M_{ts}{\rd} s\right)^+\right\vert N_t\right]\right].
\end{eqnarray}
Our task is first to compute the conditional expectation, which is essentially
Gaussian, and then the unconditional expectation, which is an expectation
over the Poisson randomness. As we have shown in the previous section,
if $\{\phi_s \}$ is decreasing for $s\geq0$ then we can find a number $\xi^*$ such
that $P(t,T,\xi^*)=K$, where
\begin{eqnarray}
P(t,T,\xi)=\frac{\int_T^\infty\rho_s \exp \left(\phi_s \xi-
\frac{1}{2}\phi_s ^2t-\lambda t({\rm e}^{\phi_s \mu+\frac{1}{2}\phi_s ^2
\delta^2}-1)\right){\rd} s} {\int_t^\infty\rho_s \exp \left(\phi_s \xi-\frac{1}{2}
\phi_s ^2t-\lambda t({\rm e}^{\phi_s \mu+\frac{1}{2}\phi_s ^2
\delta^2}-1)\right){\rd} s}.
\end{eqnarray}
Hence, after a calculation, we are able to deduce that the price of a call option in
the case of a geometric jump-diffusion martingale family is given by
\begin{eqnarray}
C_{0t} &=& \sum_{n=0}^\infty
\,\left(\int_T^\infty\frac{{\rm e}^{-\Lambda(s)t}\left(\Lambda(s)t\right)^n}{n!}\rho_s 
N\left(\frac{\xi^*-n\mu}{v_n(t)}-\phi_s v_n(t)\right){\rd} s\right. \nonumber \\ &&
\qquad\left.-K\int_t^\infty\frac{{\rm e}^{-\Lambda(s)t}\left(\Lambda(s)t\right)^n}{n!}
\rho_s N\left(\frac{\xi^*-n\mu}{v_n(t)}-\phi_s v_n(t)
\right){\rd} s\right),
\end{eqnarray}
where $\Lambda(s)=\lambda\exp\left(\phi_s \mu-\half\phi_s ^2\delta^2\right)$ 
and $v_n(t)^2=t+n\delta^2$.

\section{Geometric Gamma Family}\label{sg}
 
\noindent  We
begin with a brief review of the theory of the gamma processes. Let $\alpha$
and $\beta$ be positive constants. By a
gamma process with growth rate $\alpha$ and variance rate $\beta^2$ we mean
a process $\{\gamma_t\}_{t\geq0}$ with independent increments such that
$\gamma_0=0$ and such that $\gamma_t$ has a gamma distribution with
mean $\alpha t$ and variance $\beta^2 t$. Writing $m=\alpha^2/\beta^2$ and
$\kappa=\beta^2/\alpha$, we have $\alpha=\kappa m$ and $\beta^2=\kappa^2m$.
The density of $\gamma_t$ is then given by
\begin{eqnarray}
\mathbb{P}(\gamma_t\in{\rd} u)=\frac{u^{mt-1}{\rm e}^{-u/\kappa}}
{\kappa^{mt}\Gamma(mt)}{\rd} u
\end{eqnarray}
for $u>0$. Here $\Gamma(a)$ is the standard gamma function, which for
$a>0$ is defined by
\begin{eqnarray}
\Gamma(a)=\int_0^\infty u^{a-1}{\rm e}^{-u}{\rd} u.
\end{eqnarray}
A calculation shows that for $\lambda>-\kappa^{-1}$ the moment generating
function of $\gamma_t$ is given by
\begin{eqnarray}
\mathbb{E}\left[{\rm e}^{-\lambda\gamma_t}\right]=(1+\kappa\lambda)^{-mt},
\end{eqnarray}
from which it follows that $\mathbb{E}[\gamma_t]=\kappa mt$ and
$\text{var}[\gamma_t]=\kappa^2mt$. The exponential martingale associated
with $\{\gamma_t\}$ is given by $(1+\kappa\lambda)^{mt}
{\rm e}^{-\lambda\gamma_t}$. See Schoutens (2003), Cont \& Tankov (2004),
 Yor (2007), and Brody \textit{et al}.~(2008)  for further details of the gamma process.
 
\begin{figure}[t!]
\centering
\begin{center}
\includegraphics[width=8cm,height=10cm,keepaspectratio]{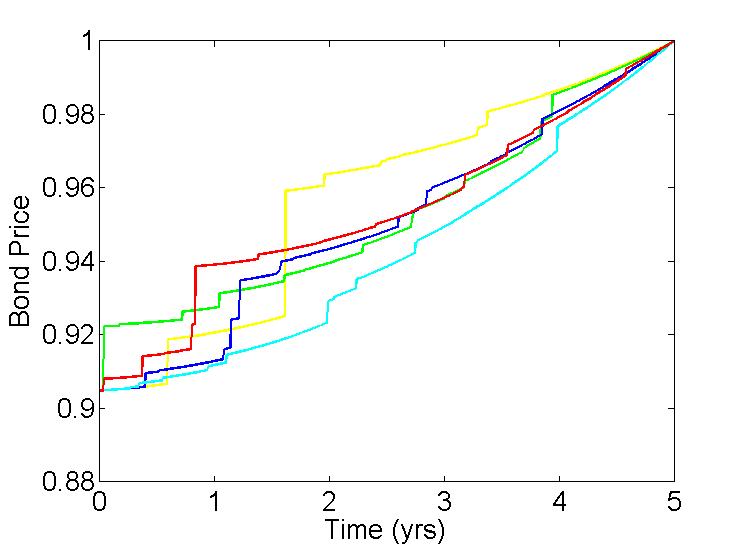}
\includegraphics[width=8cm,height=10cm,keepaspectratio]{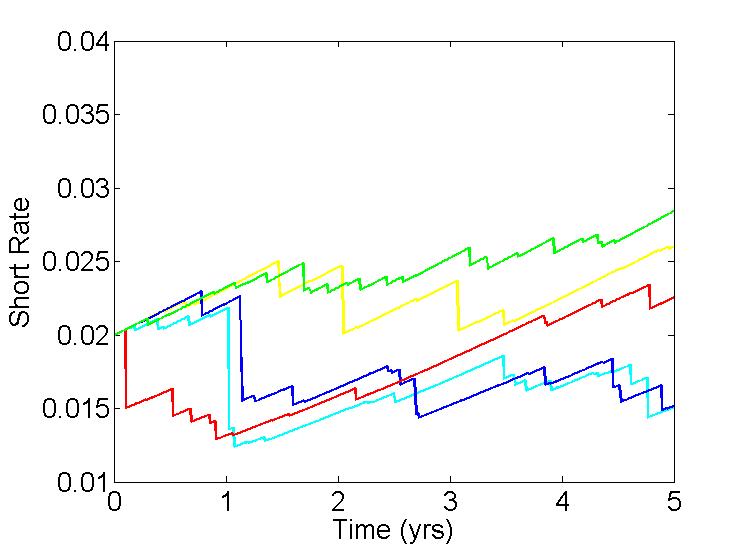}
\caption{\textit{Gamma family}
Simulations of the bond price (\ref{gbp}) and the short rate (\ref{gsr}) in
a rational term-structure model with a parametric martingale family based on a
geometric gamma process. The bond maturity is five years, the initial term
structure is given by $P_{0t}={\rm e}^{-0.02t}$, and we set
$\phi_s =-{\rm e}^{-0.02s}$. The rate parameter of the underlying gamma
process is $m=1$ and the scale parameter is $\kappa=0.5$.}\label{figg}
\end{center}
\end{figure}

Now fix $m$ and $\kappa$, let the function $\{\phi_s \}$ satisfy
$\phi_s <\kappa^{-1}$ for $s\geq0$, and define a one-parameter family of
positive martingales $\{M_{ts}\}$ by setting
\begin{eqnarray}
M_{ts}=(1-\kappa\phi_s )^{mt}{\rm e}^{\phi_s \gamma_t}.
\end{eqnarray}
Writing \{$\rho_s \}$ as before for the initial term structure density, we
obtain from equation (\ref{eq:7}) the following expression for the discount
bond prices:
\begin{eqnarray}\label{gbp}
P_{tT}=\frac{\int_T^\infty\rho_s (1-\phi_s \kappa)^{mt}{\rm e}^{\phi_s 
\gamma_t}{\rm {\rd} }s} {\int_t^\infty\rho_s (1-\phi_s \kappa)^{mt}
{\rm e}^{\phi_s \gamma_t}{\rm {\rd} }s},
\end{eqnarray}
and for the associated short rate we have
\begin{eqnarray}\label{gsr}
r_t=\frac{\rho_t (1-\phi_t \kappa)^{mt}{\rm e}^{\phi_t \gamma_t}}
{\int_t^\infty\rho_s (1-\phi_s \kappa)^{mt}{\rm e}^{\phi_s \gamma_t}{\rd} s}.
\end{eqnarray}
Sample paths associated with the bond price and the short rate are shown 
in Figure \ref{figg}. Since a gamma process only has upward jumps, and 
since the bond price is an increasing function of the underlying L\'evy 
process when $\{\phi_s\}$ is (negative and) increasing, we find that jumps in 
$\{\gamma_t\}$ induce downward jumps in the short rate process, as is 
evident from the simulation. 

In the geometric gamma model we can obtain a semi-analytical expression for
the price of a European style call option with strike price $K$. As in previous
sections we know that if $\{\phi_s \}$ is negative and increasing for $s\geq0$ 
then we can find a $\xi^*$ such that $P(t,T,\xi^*)=K$, where in this case we define
\begin{eqnarray}
P(t,T,\xi)=\frac{\int_T^\infty\rho_s (1-\phi_s \kappa)^{mt}{\rm e}^{\phi_s \xi}
{\rd} s} {\int_t^\infty\rho_s (1-\phi_s \kappa)^{mt}{\rm e}^{\phi_s \xi}{\rd} s}.
\end{eqnarray}
We deduce, using (\ref{genop}), that the price of the call option is given by
\begin{eqnarray}
C_{0t} = \int_T^\infty\rho_s \,\Gamma\left(mt,\xi^*\left({\kappa}^{-1}-
\phi_s \right)\right){\rd} s
-K\int_t^\infty\rho_s \,\Gamma\left(mt,\xi^*\left({\kappa}^{-1}-\phi_s \right)
\right){\rd} s.
\end{eqnarray}
Here we have written
\begin{eqnarray}
\Gamma(a,x)=\int_x^\infty\frac{u^{a-1}{\rm e}^{-u}}{\Gamma(a)}{\rd} u
\end{eqnarray}
for the upper incomplete gamma function.

\section{Geometric VG Family}\label{svg}

\noindent A more sophisticated model can be constructed if the underlying
L\'evy process is taken to be of the variance gamma (VG) type. The VG
process was introduced in the finance literature by Madan \& Seneta (1990),
and since then has been studied by a number of authors (see, e.g., Madan
{\em et al}.~1998). It will be useful for our purposes to begin with a brief
exposition of the theory of the VG process, treating it in a manner consistent
with our earlier discussion of the gamma process. Let $\{\gamma_{1,t}\}_{t\geq0}$ and $\{\gamma_{2,t}\}_{t\geq0}$ be a pair of
independent gamma processes, each with scale parameter unity and rate
parameter $m$. Thus $\mathbb{E}[\gamma_{1,t}]=mt$ and
$\text{var}[\gamma_{1,t}]=mt$, and similarly for $\{\gamma_{2,t}\}$. Now let
$\kappa_1$, $\kappa_2$ be a pair of nonnegative constants and set
\begin{eqnarray}
U_t=\kappa_1\gamma_{1,t}-\kappa_2\gamma_{2,t}.
\end{eqnarray}
To investigate the properties of the process thus defined we calculate the moment
generating function of $U_t$. The result is:
\begin{eqnarray}\label{mgfg}
\mathbb{E}\left[{\rm e}^{\lambda(\kappa_1\gamma_{1,t}-
\kappa_2\gamma_{2,t})}\right]=\left(1-\kappa_1
\lambda\right)^{-mt} \left(1+\kappa_1\lambda\right)^{-mt} 
= \left(1-(\kappa_1-\kappa_2)\lambda-\kappa_1 \kappa_2\lambda^2\right)^{-mt},
\end{eqnarray}
for $\lambda$ in a suitable range. We claim that $\{U_t\}$ is identical in
law to a process of the form
\begin{eqnarray}
V_t=\mu\Gamma_t+\sigma W_{\Gamma_t},
\end{eqnarray}
where $\mu$ and $\sigma$ are constants, where $\{\Gamma_t\}_{t\geq0}$ is
a scaled gamma process satisfying
$\mathbb{E}[\Gamma_t]=t$, 
and where $\{W_{\Gamma_t}\}_{t\geq0}$ represents the subordination of a
standard Brownian motion by $\{\Gamma_t\}$. We shall refer to $\{V_t\}$ as
a drifted VG process. To see the relation between $\{U_t\}$ and $\{V_t\}$ we
calculate the moment generating function of $V_t$ to obtain
\begin{eqnarray}\label{mgfb}
\mathbb{E}\left[{\rm e}^{\lambda(\mu\Gamma_t+\sigma W_{\Gamma_t})}\right]
= \left(1-\kappa_{\Gamma}\mu\lambda-\half
\kappa_{\Gamma}\sigma^2\lambda^2\right)^{-m_{\Gamma}t}.
\end{eqnarray}
\begin{figure}[t!]
\centering
\begin{center}
\includegraphics[width=8cm,height=10cm,keepaspectratio]{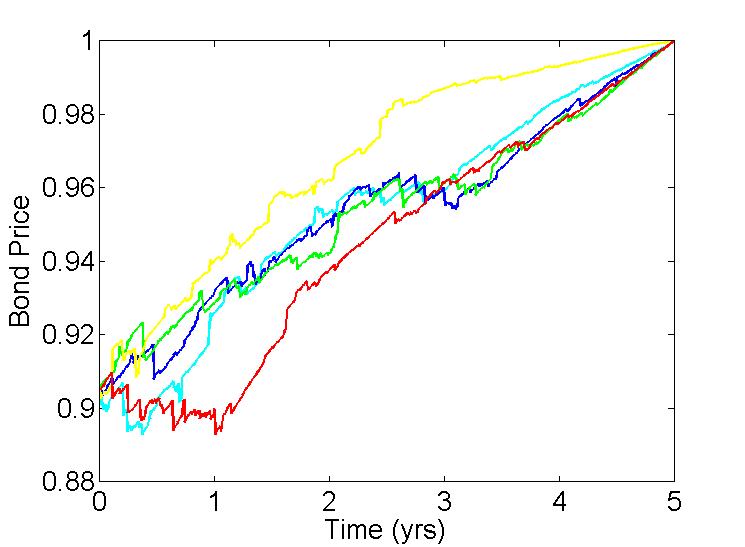}
\includegraphics[width=8cm,height=10cm,keepaspectratio]{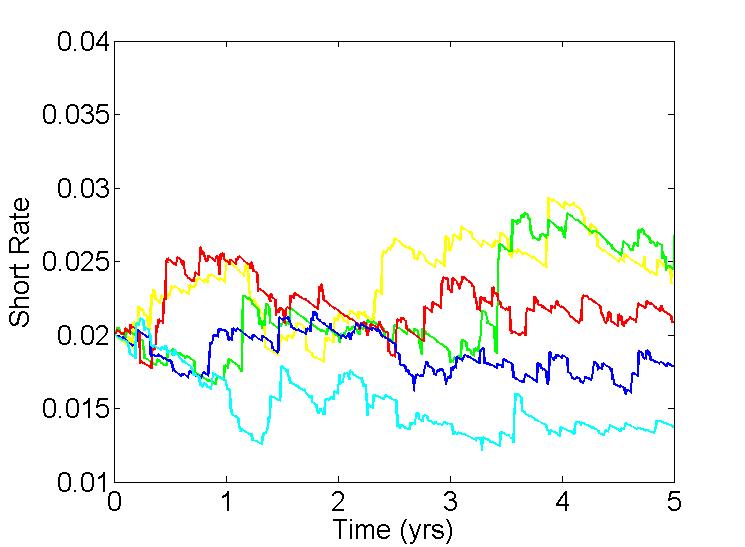}
\caption{\textit{VG family}.
Simulations of the bond price (\ref{vgbp}) and the short rate (\ref{vgsr})
in a rational term-structure model with a parametric martingale family based on
a geometric variance gamma process. The bond maturity is five years, the initial
term structure is given by $P_{0t}={\rm e}^{-0.02t}$, and we set
$\phi_s ={\rm e}^{-0.02s}$. The parameters of the VG
process are $\mu=0.5$, $\sigma=0.3$, and $m=5$.}\label{figvg}
\end{center}
\end{figure}

\noindent Here $\kappa_{\Gamma}$ is the scale parameter of $\{\Gamma_t\}$ and
$m_{\Gamma}$ is the rate parameter of $\{\Gamma_t\}$. We observe that
(\ref{mgfg}) and (\ref{mgfb}) take the same form if $m_\Gamma=m$. The
two moment generating functions can then be identified if we set
$\kappa_\Gamma\mu=\kappa_1-\kappa_2$ and
$\half\kappa_\Gamma\sigma^2=\kappa_1\kappa_2$. Next we impose the
normalisation $\mathbb{E}[\Gamma_t]=t$, which implies $\kappa_\Gamma m_\Gamma=1$.
This allows us to express $\mu$ and $\sigma$ in terms of $\kappa_1$,
$\kappa_2$, and $m$. We find that
$
\mu=(\kappa_1-\kappa_2)m $
and
$
\sigma^2=2\kappa_1\kappa_2m,
$
and hence
\begin{eqnarray}
\kappa_1=\frac{1}{2m}\left(\mu+\sqrt{\mu^2+2\sigma^2m}\right)
\quad {\rm and} \quad
\kappa_1=\frac{1}{2m}\left(-\mu+\sqrt{\mu^2+2\sigma^2m}\right).
\end{eqnarray}
Since $\{U_t\}$ and $\{V_t\}$ are L\'evy processes, the fact that the moment
generating functions agree is sufficient to ensure that the processes are
identical in law. The theory of the VG process thus outlined is consistent with that of Madan
{\em et al}.~(1998). The parametrisation that we have chosen is for our purposes
more transparent. In particular, the limiting cases where
$\kappa_1=0$ or $\kappa_2=0$ are incorporated. We 
define a family of positive martingales by
setting
\begin{eqnarray}\label{vgmart}
M_{ts}=\left(1-\frac{\mu}{m}\phi_s -\frac{\sigma^2}{2m}\phi_s ^2\right)^{mt}
{\rm e}^{\phi_s V_t}.
\end{eqnarray}
Then, using equation (\ref{eq:7}), we deduce that the bond price takes the form
\begin{eqnarray}\label{vgbp}
P_{tT}=\frac{\int_T^\infty\rho_s (1-\frac{\mu}{m}\phi_s -\frac{\sigma^2}{2m}
\phi_s ^2)^{mt}{\rm e}^{\phi_s V_t}{\rd} s} {\int_t^\infty\rho_s 
(1-\frac{\mu}{m}\phi_s -\frac{\sigma^2}{2m}\phi_s ^2)^{mt}
{\rm e}^{\phi_s V_t}{\rd} s},
\end{eqnarray}
and that the associated short rate is given by
\begin{eqnarray}\label{vgsr}
r_t=\frac{\rho_t (1-\frac{\mu}{m}\phi_t -\frac{\sigma^2}{2m}
\phi_t ^2)^{mt}{\rm e}^{\phi_t V_t}} {\int_t^\infty
\rho_s (1-\frac{\mu}{m}\phi_s -\frac{\sigma^2}{2m}
\phi_s ^2)^{mt}{\rm e}^{\phi_s V_t}{\rd} s}.
\end{eqnarray}
Simulations of the bond price (\ref{vgbp}) and the short rate (\ref{vgsr}) are
presented in Figure \ref{figvg}.

We proceed to derive the price of a bond option in the setting of the geometric
VG martingale family. In the case of a call option with expiry $t$ and strike $K$
on a bond with maturity $T$, let us consider expression (\ref{optpr}), into which
we substitute (\ref{vgmart}).
As before, when $\{\phi_s \}$ is decreasing for $s\geq0$ we are able to find
a number $\xi^*$ such that $P(t,T,\xi^*)=K$, where
\begin{eqnarray}
P(t,T,\xi)=\frac{\int_T^\infty\rho_s (1-\frac{\mu}{m}\phi_s -\frac{\sigma^2}{2m}
\phi_s ^2)^{mt}{\rm e}^{\phi_s \xi}
{\rd} s} {\int_t^\infty\rho_s (1-\frac{\mu}{m}\phi_s -\frac{\sigma^2}{2m}
\phi_s ^2)^{mt}{\rm e}^{\phi_s \xi}{\rd} s}.
\end{eqnarray}
In terms of the critical level $\xi^*$ we
deduce that the price of the option is given by
\begin{eqnarray}
C_{0t}&=&
\int_T^\infty\rho_s \Psi\left[\frac{\xi^*}{\sigma\Phi(s) },-
\left(\frac{\mu}{\sigma}+\sigma\phi_s \right)\Phi(s) ,mt\right]{\rd} s
\nonumber \\ &&
-K\int_t^\infty\rho_s \Psi\left[\frac{\xi^*}{\sigma \Phi(s) },
-\left(\frac{\mu}{\sigma}+\sigma\phi_s \right)\Phi(s)  ,mt\right]{\rd} s,
\end{eqnarray}
where
$\Phi(s) =\left(m-\mu\phi_s -\half
\sigma^2\phi_s ^2\right)^{-\half}$
and
\begin{eqnarray}\label{Psi}
\Psi(a,b,c)=\int_0^\infty N \left(\frac{a}{\sqrt{u}}+b\sqrt{u}\right)
\frac{u^{c-1}{\rm e}^{-u}}{\Gamma(c)}{\rd} u.
\end{eqnarray}

\begin{figure}[t!]
\centering
\begin{center}
\includegraphics[width=8cm,height=10cm,keepaspectratio]{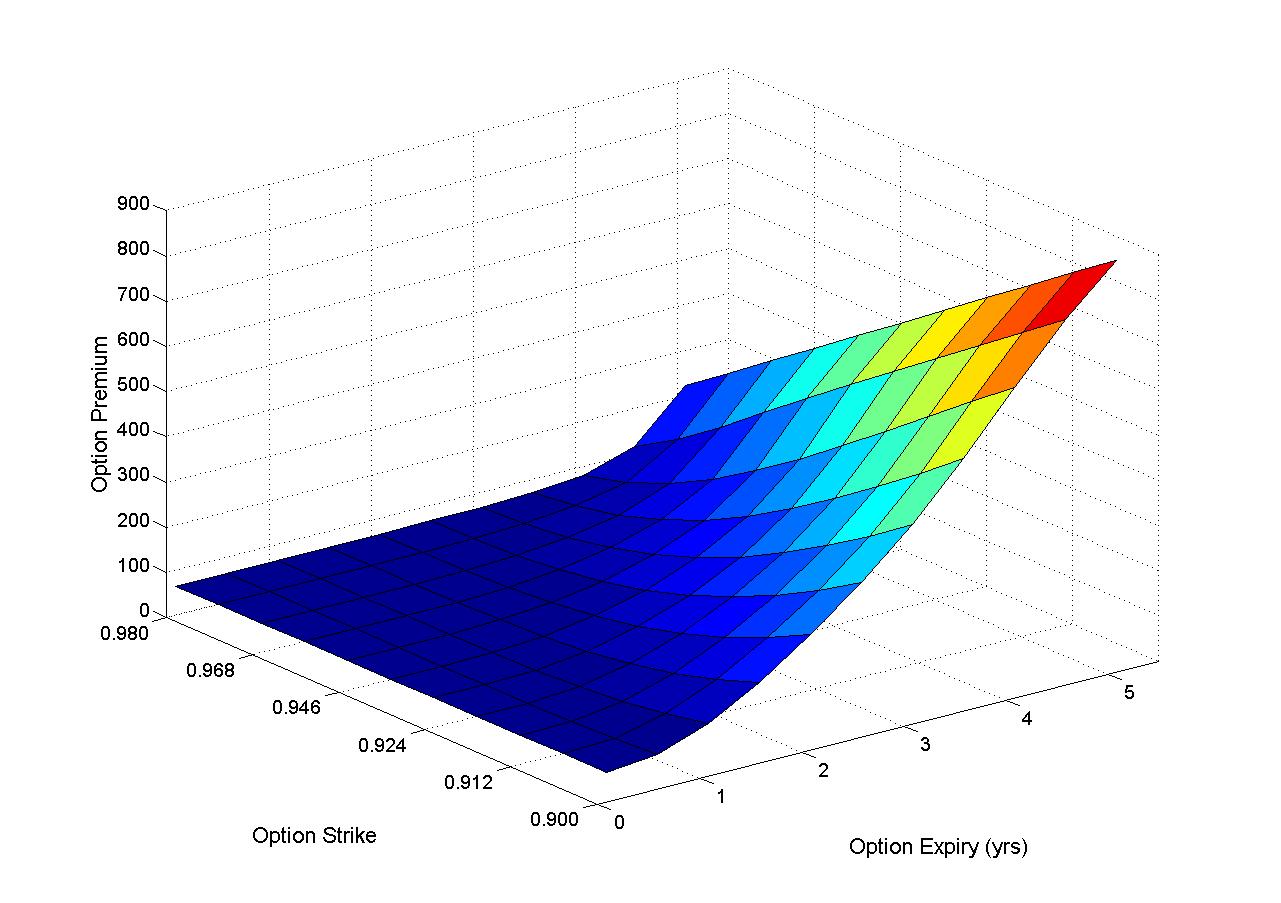}
\includegraphics[width=8cm,height=10cm,keepaspectratio]{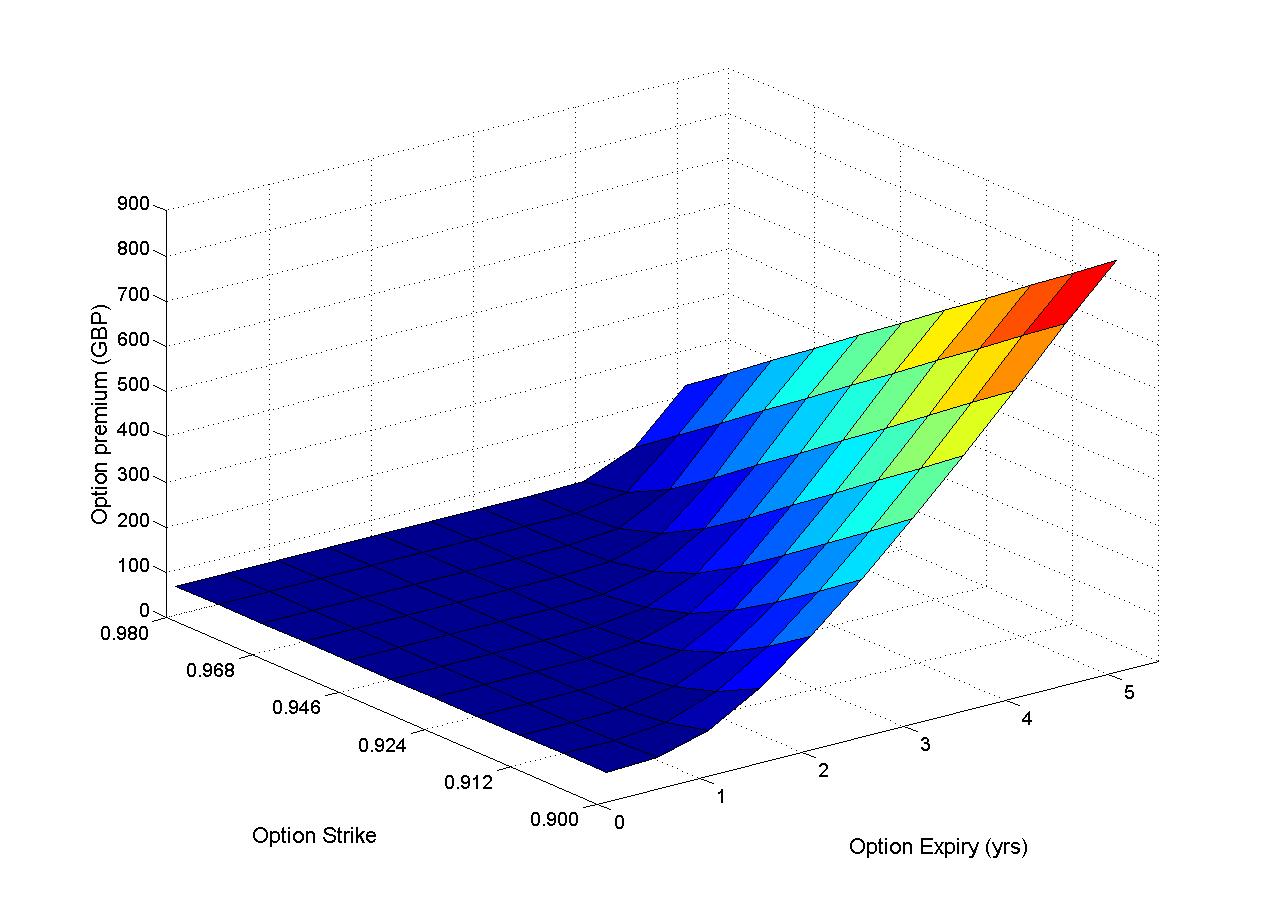}
\caption{\textit{Option surfaces}. 
Simulations of the option price (\ref{eq:OptionPrice}) when the underlying 
term-structure model is based on (left) the geometric jump-diffusion process 
(\ref{eq:GeoMartJD}) and (right) the geometric variance gamma process 
(\ref{vgmart}). The initial term structure is given by $P_{0t}={\rm e}^{-0.03t}$, and we 
set $\phi_s ={\rm e}^{-0.02s}$. The rate parameter of the Poisson process is 
$\lambda=5$. The jump sizes have mean $\mu=0$, and standard deviation 
$\delta=1$. The parameters of the VG
process are $\mu=0.02$, $\sigma=0.3$, and $m=20$.}\label{fig:OptSurfaces}
\end{center}
\end{figure}

In Figure \ref{fig:OptSurfaces} 
we present examples of price surfaces of European-style call options for the 
jump-diffusion and VG families of models. The figures each contain one 
hundred prices computed over a range of strikes and option expiries. The 
computing times for one-hundred call prices for the Gaussian, jump-diffusion, 
gamma, and VG families of term-structure models were 6 seconds, 54 
seconds, 10 seconds, and 6 seconds, respectively.

\begin{acknowledgments}
\noindent The authors would like to thank seminar participants at the Imperial College
Workshop on Stochastics, Control and Finance, London, April 2010, at the
Fifth General Conference on Advanced Mathematical Methods in Finance, Bled,
Slovenia, May 2010, at the Fields Institute Workshop on
Financial Derivatives and Risk Management, Toronto, May 2010, at the
Sixth World Congress of the Bachelier Finance Society, Toronto, June 2010,
at the Workshop on Mathematical Finance and Related Issues, Kyoto,
September 2010, at the Bank of Japan, Tokyo, September 2010, and
at the Department of Economics, Hitotsubashi University, Tokyo,
September 2010, for useful comments. L.~P.~Hughston thanks R.~Miura,
J.~Sekine, and H.~Sugita for hospitality.
E.~Mackie acknowledges support by EPSRC, and
thanks A.~S.~Iqbal and S.~Lyons for helpful discussions.
\end{acknowledgments}
%

\vskip 15pt 


\end{document}